\documentclass[a4paper,11pt]{article}

\usepackage{amssymb}
\usepackage{amsmath}
\usepackage{graphicx}
\usepackage{comment}

\newcommand{\bin}[2]{\left(\begin{array}{c} \!\!#1\!\! \\  \!\!#2\!\! \end{array}\right)}
\newcommand{\troisj}[3]{\left(\begin{array}{ccc}#1 & #2 & #3 \\ 0 & 0 & 0 \end{array}\right)}
\newcommand{\troisjm}[6]{\left(\begin{array}{ccc}#1 & #2 & #3 \\ #4 & #5 & #6 \end{array}\right)}

\begin{document}

\huge

\begin{center}
Koopmans' theorem in statistical Hartree-Fock theory
\end{center}

\vspace{1cm}

\large

\begin{center}
Jean-Christophe Pain\footnote{jean-christophe.pain@cea.fr}
\end{center}

\normalsize

\begin{center}
CEA, DAM, DIF, F-91297 Arpajon, France\\
\end{center}

\vspace{1cm}

\begin{abstract}
In this short paper, the validity of Koopmans' theorem in the Hartree-Fock theory at non-zero temperature (Hartree-Fock statistical theory) is investigated. It is shown that Koopmans' theorem does not apply in the grand-canonical ensemble, due to a missing contribution to the energy proportional to the interaction between two electrons belonging to the same orbital. Hartree-Fock statistical theory has also been applied in the canonical ensemble [Blenski \emph{et al.}, Phys. Rev. E \textbf{55}, R4889 (1997)] for the purpose of photo-absorption calculations. In that case, the Hartree-Fock self-consistent-field equations are derived in the super-configuration approximation. It is shown that Koopmans' theorem does not hold in the canonical ensemble, but that a restricted version of the theorem can be obtained, by assuming that a particular quantity multiplying the interaction matrix element in the expression of the energy does not change during the removal of an electron.
\end{abstract}

\section{Introduction}

In atomic physics of hot dense plasmas, the Hartree-Fock (HF) theory is applied to model self-consistent potentials of atoms and ions. Another interest of HF theory is to use HF one-electron states in order to calculate the average configuration energies. As shown by Mermin \cite{MERMIN63}, the HF theory at non-zero temperature is equivalent to minimizing the grand thermodynamic potential in a restricted class of statistical operators. In the following, such a theory will be refered to as ``grand-canonical statistical HF theory''.

A few years ago, Blenski \emph{et al.} \cite{BLENSKI97a,BLENSKI97b} generalized the statistical Hartree-Fock approach to atoms in plasmas at finite temperature in the framework of the super-configuration approximation successfully applied in the STA (Super Transition Array) method \cite{BAR89}. In the following, such a theory will be refered to as ``canonical statistical HF theory''.

Koopmans' theorem \cite{KOOPMANS33,FROESE77,GRANT07} states that, when the number of electrons $q_i$ in a given orbital $i$ is greater than 1, the interaction within the group is generally much smaller than the energy difference resulting from the removal of an electron. Therefore, $\epsilon_i$ represents roughly the average theoretical ionization potential for a single electron during the removal of the group. Koopmans \cite{KOOPMANS33} identified the physically meaningful solutions as those for which $\epsilon_i$ was still an approximation to the ionization potential when $q_i=1$, and an approximation to the average ionization potential otherwise. An important point is that Koopmans' theorem is valid only if the one-electron wavefunctions in the $N$-electron and the ($N\pm$1)-electron Slater determinants are the same. 

The purpose of this short paper is to check whether Koopmans' theorem is valid in the framework of statistical HF theory in the grand-canonical ensemble \cite{MERMIN63} and in the canonical ensemble \cite{BLENSKI97a,BLENSKI97b}. For that purpose, the energy of the system with $q_i$ electrons in orbital $i$ is compared to the one with $q_i-1$ electrons in orbital $i$. The difference is related to the value of the energy of orbital $i$.

\section{Grand-canonical ensemble}

\subsection{Average energies}

The total energy of a configuration with $N$ orbitals $(n_1l_1),~(n_2l_2),~\cdots,~(n_Nl_N)$ is \cite{SLATER60}

\begin{equation}
E=\sum_{i=1}^Nq_i~I_i+\frac{1}{2}\sum_{i,j=1}^Nq_i\left(q_j-\delta_{i,j}\right)V_{ij}
\end{equation}

where $q_i$ is the population (number of electrons) of orbital $i$ and, in the non-relativistic case (Schr\"{o}dinger equation), the one-electron energy $I_i$ reads (atomic units are used throughout the article):

\begin{equation}
I_i=\int_0^{\infty}P_i(r)\left(-\frac{1}{2}\left[\frac{d^2}{dr^2}-\frac{l(l+1)}{r^2}\right]-\frac{Z}{r}\right)P_i(r)~dr,
\end{equation}

where $P_i(r)$ is the radial part of the wavefunction multiplied by $r$. The interaction matrix elements are

\begin{equation}
V_{ii}=F^{(0)}(ii)-\frac{2l_i+1}{4l_i+1}\sum_{k\ne 0}\troisj{l_i}{k}{l_i}^2 F^{(k)}(ii)
\end{equation}

and for $i\ne j$,

\begin{equation}
V_{ij}=F^{(0)}(ij)-\frac{1}{2}\sum_{k\ne 0}\troisj{l_i}{k}{l_j}^2 G^{(k)}(ii),
\end{equation}

where $F^{(k)}$ and $G^{(k)}$ are the direct and exchange Slater integrals respectively \cite{COWAN81}. The quantity $\troisjm{j_1}{j_2}{j_3}{m_1}{m_2}{m_3}$ represents a $3j$-symbol \cite{EDMONDS57}. The average energy of the configuration is

\begin{eqnarray}\label{ave1}
\langle E\rangle&=&\frac{1}{\mathcal{Z}}\sum_{q_1=0}^{g_1}\sum_{q_2=0}^{g_2}\sum_{q_3=0}^{g_3}\cdots\sum_{q_N=0}^{g_N}\left[\sum_{i=1}^Nq_i~I_i+\frac{1}{2}\sum_{i,j=1}^Nq_i\left(q_j-\delta_{i,j}\right)V_{ij}\right] \nonumber\\
&&\times\prod_{k=1}^N\bin{g_k}{q_k}e^{-\beta\left(\epsilon_k-\mu\right)~q_k}
\end{eqnarray}

where $\bin{a}{b}=a!/(b!(a-b)!)$ is the binomial coefficient, $\mu$ the chemical potential and $\mathcal{Z}$ the partition function of the system:

\begin{equation}
\mathcal{Z}=\sum_{q_1=0}^{g_1}\sum_{q_2=0}^{g_2}\sum_{q_3=0}^{g_3}\cdots\sum_{q_N=0}^{g_N}\prod_{k=1}^N\bin{g_k}{q_k}e^{-\beta\left(\epsilon_k-\mu\right)~q_k}.
\end{equation}

One finds that the average energy (\ref{ave1}) can be written

\begin{eqnarray}
\langle E\rangle&=&=\sum_{i=1}^N\langle q_i\rangle I_i+\frac{1}{2}\sum_{i,j=1}^N\langle q_i\rangle\left(\langle q_j\rangle-\frac{\langle q_i\rangle}{g_i}~\delta_{i,j}\right)V_{ij}\nonumber\\
&=&\sum_{i=1}^N\langle q_i\rangle I_i+\frac{1}{2}\sum_{i,j=1}^N\langle q_i\rangle\left(\langle q_j\rangle-p_i~\delta_{i,j}\right)V_{ij},
\end{eqnarray}

with

\begin{equation}
\langle q_i\rangle=g_i~p_i=\frac{g_i}{1+e^{-\beta\left(\epsilon_i-\mu\right)}}.
\end{equation}

The energy required in order to remove one electron from orbital $i$ is defined by

\begin{equation}
\Delta E^{(i)}=E\left(\{q_k-\delta_{ik}, k=1, N\})-E(\{q_k, k=1, N\}\right),
\end{equation}

where $\delta_{ik}$ denotes Kronecker's symbol. The notation $\{q_k-\delta_{ik}, k=1, N\}$ means that the population of orbital $i$ is reduced by 1, the other populations remaining unchanged. Therefore, one obtains

\begin{equation}
\Delta E^{(i)}=-I_i-\left(\langle q_i\rangle-\frac{1}{2}\right)\left(1-\frac{1}{g_i}\right)V_{ii}-\sum_{j\ne i}\langle q_j\rangle V_{ij}.
\end{equation}

\subsection{Expression of orbital energies}

The Hartree-Fock equations read \cite{FROESE77}

\begin{equation}\label{hfeq}
\left(-\frac{1}{2}\left[\frac{d^2}{dr^2}-\frac{l(l+1)}{r^2}\right]-\frac{Z}{r}+V(r)+\delta V_i(r)-\epsilon_i\right)P_i(r)+g_i(r)=0,
\end{equation}

with

\begin{equation}
V(r)=\sum_{j=1}^N\langle q_j\rangle\int_0^{\infty}P_j(r')\frac{1}{r_>}P_j(r')~dr'.
\end{equation}

The correction $\delta V_i$ is given by 

\begin{equation}
\delta V_i(r)=-p_iV^{(0)}(i,i)-\frac{2l_i+1}{4l_i+1}\sum_{k\ne 0}\troisj{l_i}{k}{l_i}^2V^{(k)}(ii)\left(\langle q_i\rangle-p_i\right),
\end{equation}

where

\begin{equation}
V^{(k)}(ij)=\int_0^{\infty}P_j(r')\frac{r_<^k}{r_>^{k+1}}P_i(r')~dr'
\end{equation}

and

\begin{equation}
g_i(r)=-\sum_{j\ne i}\frac{\langle q_j\rangle}{2}\sum_k\troisj{l_i}{k}{l_j}^2V^{(k)}(ij)~P_j(r).
\end{equation}

Multiplying equation (\ref{hfeq}) by $P_j(r)$ and integrating from 0 to $\infty$ leads to

\begin{equation}
\epsilon_i=I_i+\langle q_i\rangle\left(1-\frac{1}{g_i}\right)V_{ii}+\sum_{j\ne i}\langle q_j\rangle V_{ij}.
\end{equation}

Therefore, one has

\begin{equation}
\Delta E^{(i)}=-\epsilon_i+\frac{1}{2}\left(1-\frac{1}{g_i}\right)V_{ii}\ne-\epsilon_i,
\end{equation}

which means that Koopmans' theorem is not verified in that case, because of the remaining term $(1-1/g_i)V_{ii}/2$. Each orbital is shifted from that quantity, which is the signature of the competition between the diagonal term of the interaction matrix and the degeneracy. We can see in table \ref{tab1}, in the case of a carbon plasma at $T$=30 eV and $\rho$=0.01 g/cm$^3$, that the difference between the energy difference $\Delta E^{(i)}$ and $\epsilon_i$ can reach 25 $\%$. The shift is more pronounced for the higher-energy orbitals, which can be explained by the fact that the electrons in such orbitals are very sensitive to electron-electron interactions, unlike electrons in the lower orbitals which are more subject to the attraction of the nucleus\footnote{However, it is worth mentioning that Janak's theorem \cite{JANAK78,WILSON95} holds in that case, \emph{i.e.}

\begin{equation}
\frac{\partial\langle E\rangle}{\partial\langle q_i\rangle}=\epsilon_i.
\end{equation}}.

\begin{table}[t]
\begin{center}
\begin{tabular}{|c|c|c|c|c|c|c|} \hline\hline
Orbital & $g_i$ & {\bf Energy} $\bf\epsilon_i$ (eV) & $V_{ii}$ & $(1-1/g_i)V_{ii}/2$ & $\bf\Delta E^{(i)}$ (eV) & Shift ($\%$)\\ \hline\hline
$1s$ & 2 & -383.376 & 44.892 & 11.223 & 394.599 & 2.93\\
$2s$ & 2 & -57.101 & 9.212 & 2.303 & 59.404 & 4.03\\
$2p$ & 6 & -48.905 & 15.994 & 6.664 & 55.569 & 13.63\\
$3s$ & 2 & -17.519 & 3.791 & 0.948 & 18.467 & 5.41\\
$3p$ & 6 & -16.996 & 6.312 & 2.630 & 19.626 & 15.47\\
$3d$ & 10 & -16.188 & 8.088 & 3.640 & 19.828 & 22.48\\
$4s$ & 2 & -6.934 & 2.062 & 0.515 & 7.449 & 7.43\\ 
$4p$ & 6 & -6.154 & 3.749 & 1.562 & 7.716 & 25.38\\\hline\hline
\end{tabular}
\caption{Energies (in eV) of orbitals $1s$ to $4p$ for a carbon plasma at $T$=30 eV and $\rho$=0.01 g/cm$^3$. The chemical potential is $\mu$=-186.702 eV.}\label{tab1}
\end{center}
\end{table}

\section{Canonical ensemble}

The formalism presented in the preceeding section allows non-integer populations for the different orbitals. Therefore, it provides the average (in the sense of the most probable) configuration of the plasma. From this average configuration, the real configurations can be built, by rounding the population values to the closest integer. The problem is that, in hot plasmas, the number of configurations can be really huge. The super-configuration method \cite{BAR89} has been invented in order to remedy this problem. A configuration is made of orbitals with integer populations; in the same way, a super-configuration is made of super-orbitals with integer populations, a super-orbital being a group of orbitals which energies are close to each other. For instance, $(1s2s2p)^3(3s3p)^4(3d)^7$ is a super-configuration made of 3 super-orbitals populated respectively with 3, 4 and 7 electrons. 

The super-configuration approximation enables one to calculate the equation of state beyond the average-atom model \cite{PAIN02,PAIN06,PAIN07}. The idea is to study the influence of the population fluctuations on the thermodynamic quantities. For instance, the pressure of the plasma is given by

\begin{equation}
P=\sum_{\Xi}W_{\Xi}~P_{\Xi},
\end{equation}

where $W_{\Xi}$ and $P_{\Xi}$ represent respectively the probability and the pressure of super-configuration $\Xi$. In the particular case where a super-configuration is an ordinary configuration, the equations reduce to the standard HF equations. The finite-temperature HF method was derived in the framework of the super-configuration approximation by Blenski \emph{et al.} \cite{BLENSKI97a,BLENSKI97b}. The average shell populations and interaction matrices, which are averages of the corresponding quantities for configurations, are given in terms of partition functions \cite{GILLERON04,WILSON07,PAIN09}. This allows one to avoid problems stemming from non-integer population numbers in other thermal HF theories \cite{WILSON95}.
 
\subsection{Expression of the average energy of a $Q$-electron configuration}

The Hartree-Fock equations read \cite{BLENSKI97a,BLENSKI97b}:

\begin{eqnarray}\label{hfeq2}
& &\left(-\frac{1}{2}\left[\frac{d^2}{dr^2}-\frac{l(l+1)}{r^2}\right]-\frac{Z}{r}+V(r)+\delta V_i(r)-\epsilon_i\right)P_i(r)+h_i(r)\nonumber\\
&=&\sum_{k=1, k\ne i}^N\epsilon_{i,k}~\langle q_k\rangle_Q~\delta_{l_i,l_k}~P_k(r).
\end{eqnarray}

The different terms in the left-hand-side of Eq. (\ref{hfeq}) are

\begin{eqnarray}
\delta V_i(r)&=&\sum_{s,k=1}^N\left\{\delta_{i,s}\left(g_i-1\right)\left[g_i\left(S_i-p_i\right)\delta_{k,0}-S_i\right]\right.\nonumber\\
& &\left.+\left(1-\delta_{i,s}\right)\langle q_s\rangle_Q\left(g_s-1\right)H_{is}~\delta_{k,0}\right\}Y_{s,s}^{(k)}(r),
\end{eqnarray}

and

\begin{equation}
h_i(r)=-\sum_{s=1,s\ne i}^N\langle q_s\rangle_Q\left[H_{is}+1\right]P_s(r)\sum_{k=1}^N Y_{s,i}^{(k)}(r),
\end{equation}

with

\begin{equation}
Y_{s,i}^{(k)}(r)=\frac{g_i}{2(g_i-\delta_{s,i})}\troisj{l_i}{k}{l_s}^2\int_0^{\infty}\frac{r_<^k}{r_>^{k+1}}P_s(r')P_i(r')~dr'.
\end{equation}

There is a small typographical error in Ref. \cite{BLENSKI97a}: a factor ($g_s-$1) is missing in Eq. (6c). However, in Eq. (7c) of Ref. \cite{BLENSKI97b}, the expression is correct. Multiplying equation (\ref{hfeq2}) by $P_j(r)$ and integrating over $r$ from 0 to infinity leads to the orbital energy

\begin{equation}
\epsilon_i=I_i+S_i\left(g_i-1\right)V_{ii}+\sum_{s\ne i}\left[1+H_{is}\right]V_{is}\langle q_s\rangle_Q.
\end{equation}

Let us consider the following super-configuration with $N$ orbitals and $Q$ electrons: 

\begin{equation}
(n_1l_1~n_2l_2~\cdots~n_Nl_N)^Q. 
\end{equation}

Its average energy reads 

\begin{equation}
\langle E\rangle_Q=\sum_{i=1}^N\langle q_i\rangle_Q~I_i+\frac{1}{2}\sum_{i,j=1}^N\langle q_i\rangle_Q\left(\langle q_j\rangle_Q-p_i~\delta_{i,j}\right)W_{ij}~V_{ij},
\end{equation}

\noindent where

\begin{equation}
\langle q_i\rangle_Q=g_i~p_i=-g_i\sum_{n=1}^Q\left(-X_i\right)^n\frac{U_{Q-n}(g)}{U_Q(g)}
\end{equation}

is the average population of orbital $i$, $X_i=e^{-\beta(\epsilon_i-\mu)}$, and 

\begin{equation}
U_Q(g)=\underbrace{\sum_{q_1=0}^{g_1}\sum_{q_2=0}^{g_2}\sum_{q_3=0}^{g_3}\cdots\sum_{q_N=0}^{g_N}}_{\sum_{j=1}^Nq_j=Q}\prod_{k=1}^N\bin{g_k}{q_k}X_k^{q_k}.
\end{equation}

One has also

\begin{equation}
W_{rs}=1+\delta_{r,s}\left[\frac{S_r\left(g_r-1\right)}{\langle q_r\rangle_Q-1}-1\right]+\left(1-\delta_{r,s}\right)H_{rs},
\end{equation}

where
 
\begin{equation}
S_i=\frac{1}{p_i}\sum_{n=1}^Q(n-1)\left(-X_i\right)^n\frac{U_{Q-n}(g)}{U_Q(g)}
\end{equation}

and

\begin{equation}
H_{rs}=\frac{1}{X_s-X_r}\left[\frac{g_s~X_s}{\langle q_s\rangle_Q}-\frac{g_r~X_r}{\langle q_r\rangle_Q}\right]-1.
\end{equation}

The energy required in order to remove one electron from orbital $i$ is

\begin{equation}
\Delta \langle E^{(i)}\rangle_Q=E\left(\left\{\langle q_k\rangle_Q-\delta_{ik}, k=1, N\right\}\right)-E\left(\left\{\langle q_k\rangle_Q, k=1, N\right\}\right)\ne-\epsilon_i,
\end{equation}

which means that Koopmans' theorem does not hold in the statistical Hartree-Fock theory for the canonical thermodynamic description of the system. If $W_{ij}$ did not change during the removal of an electron belonging to orbital $i$, one would have

\begin{equation}
\Delta E^{(i)}-=I_i-S_i\left(g_i-1\right)V_{ii}-\sum_{s\ne i}\left[1+H_{is}\right]V_{is}\langle q_s\rangle_Q=-\epsilon_i,
\end{equation}

but this is not true in general.

\section{Conclusion}

Koopmans' theorem does not apply in the statistical Hartree-Fock theory neither in the canonial ensemble, nor in the grand-canonical ensemble. In the grand-canonical ensemble, an additional term exists in the energy variation due to the removal of an electron from orbital $i$. It represents a shift of the orbital energy depending on its degeneracy and on the diagonal matrix element describing the interaction between two electrons in that orbital. In the canonical ensemble, one finds, in the framework of the super-configuration method, that Koopmans' theorem does not hold \emph{stricto sensu}, unless a particular quantity is ensured to be unchanged when the number of electrons is decreased by one.

\end{document}